\documentclass[sigconf]{acmart}
\usepackage{amsmath}
\usepackage{booktabs,caption}
\usepackage{soul}

\AtBeginDocument{%
  \providecommand\BibTeX{{%
    \normalfont B\kern-0.5em{\scshape i\kern-0.25em b}\kern-0.8em\TeX}}}

\setcopyright{acmcopyright}
\copyrightyear{2018}
\acmYear{2018}
\acmDOI{10.1145/1122445.1122456}

\acmConference[WebSci'20]{WebSci'20: The 12th International ACM Conference on Web Science 2020}{July 07-10, 2020}{Southampton, UK}

\acmISBN{978-1-4503-XXXX-X/18/06}

\begin{document}

\copyrightyear{2020}
\acmYear{2020}
\setcopyright{acmlicensed}\acmConference[WebSci '20]{12th ACM Conference on Web Science}{July 6--10, 2020}{Southampton, United Kingdom}
\acmBooktitle{12th ACM Conference on Web Science (WebSci '20), July 6--10, 2020, Southampton, United Kingdom}
\acmPrice{15.00}
\acmDOI{10.1145/3394231.3397906}
\acmISBN{978-1-4503-7989-2/20/07}

%%
%% The "title" command has an optional parameter,
%% allowing the author to define a "short title" to be used in page headers.
\title{Examining the Role of Mood Patterns in Predicting Self-Reported Depressive symptoms}

%%
%% The "author" command and its associated commands are used to define
%% the authors and their affiliations.
%% Of note is the shared affiliation of the first two authors, and the
%% "authornote" and "authornotemark" commands
%% used to denote shared contribution to the research.
\author{Lucia Lushi Chen}
\email{Lushi.Chen@ed.ac.uk}

\affiliation{%
  \institution{School of Informatics, The University of Edinburgh}
  \streetaddress{10 Crichton St}
  \city{Edinburgh}
  \country{United Kingdom}
  \postcode{EH8 9AB}
}

\author{Walid Magdy}
\email{wmagdy@inf.ed.ac.uk}

\affiliation{%
  \institution{School of Informatics, The University of Edinburgh}
  \streetaddress{10 Crichton St}
  \city{Edinburgh}
  \country{United Kingdom}
  \postcode{EH8 9AB}
}

\author{Heather Whalley}
\email{heather.whalley@ed.ac.uk}

\affiliation{%
  \institution{Centre for Clinical Brain Sciences, The University of Edinburgh}
  \streetaddress{Edinburgh BioQuarter, 49 Little France Crescent}
  \city{Edinburgh}
  \country{United Kingdom}
  \postcode{EH8 9AB}
}

\author{Maria Wolters}
\email{maria.wolters@ed.ac.uk}

\affiliation{%
  \institution{School of Informatics, The University of Edinburgh}
  \streetaddress{10 Crichton St}
  \city{Edinburgh}
  \country{United Kingdom}
  \postcode{EH8 9AB}
}

%%
%% By default, the full list of authors will be used in the page
%% headers. Often, this list is too long, and will overlap
%% other information printed in the page headers. This command allows
%% the author to define a more concise list
%% of authors' names for this purpose.
\renewcommand{\shortauthors}{Chen, et al.}

%%
%% The abstract is a short summary of the work to be presented in the
%% article.
\begin{abstract}
%Researchers have explored automatic screening models as a quick way to identify potential risks of developing depressive symptoms. In this paper, 
%We examined the potential of leveraging mood derived from social media text to infer depression symptoms. We introduced several mood representations that explicitly models the changes of mood, and transitions between moods. 

Researchers have explored automatic screening models as a quick way to identify potential risks of developing depressive symptoms. Most existing models include a person's mood as reflected on social media at a single point in time as one of the predictive variables. In this paper, we study the changes and transition in mood reflected on social media text over a period of one year using a \emph{mood profile}.  We used a subset of the "MyPersonality" Facebook data set that comprises users who have consented to and completed an assessment of depressive symptoms. The subset consists of 93,378 Facebook posts from 781 users. We observed less evidence of mood fluctuation expressed in social media text from those with low symptom measures compared to others with high symptom scores. Next, we leveraged a daily mood representation in Hidden Markov Models to determine its associations with subsequent self-reported symptoms. We found that individuals who have specific mood patterns are highly likely to have reported high depressive symptoms. However, not all of the high symptoms individuals necessarily displayed this characteristic, which indicates presence of potential subgroups driving these findings. Finally, we leveraged multiple mood representations to characterize levels of depressive symptoms with a logistic regression model. Our findings support the claim that for  some people, derived mood from social media text can be a proxy of real-life mood, in particular depressive symptoms. Combining the mood representations with other proxy signals can potentially advance responsibly used semi-automatic screening procedures. 

%Our study demonstrates the utility of time dependent social media text in inferring potential depressive symptoms.

\end{abstract}

%%
%% The code below is generated by the tool at http://dl.acm.org/ccs.cfm.
%% Please copy and paste the code instead of the example below.
%%

% \ccsdesc[300]{Computer systems organization~Redundancy}
% \ccsdesc[500]{Computing methodologies~Gaussian processes}
% \ccsdesc[600]{Computing methodologies~Machine learning algorithms}
%%
%% Keywords. The author(s) should pick words that accurately describe
%% the work being presented. Separate the keywords with commas.

%% A "teaser" image appears between the author and affiliation
%% information and the body of the document, and typically spans the
%% page.

%%
%% This command processes the author and affiliation and title
%% information and builds the first part of the formatted document.
 \maketitle

\section{Introduction}
%Social media data provides opportunities for identifying mental health issues. Numerous computational models have successfully identified mental disorder symptoms or conditions from automatic screening technologies \citep{de2013predicting,coppersmith2014quantifying,tsugawa2015recognizing,reece2017forecasting}. 
Depression is the leading cause of disability worldwide. Initial efforts to detect depression signals from social media posts have shown promising results \citep{de2013predicting,coppersmith2014quantifying,park2012depressive,tsugawa2015recognizing, nguyen2014affective, nadeem2016identifying, almeida2017detecting}. Given the high internal validity \citep{reece2017forecasting, de2013predicting}, results from such analyses are potentially beneficial to clinical judgement. The existing models for automatic detection of depressive symptoms learn proxy diagnostic signals from social media data, such as help-seeking behaviour for mental health or medication names \cite{de2013predicting, coppersmith2014quantifying}. However, in reality, individuals with depression typically experience depressed mood, loss of pleasure nearly in all the activities, feeling of worthlessness or guilt, and diminished ability to think \citep{american2013diagnostic}. Therefore, a lot of the proxy signals used in these models lack the theoretical underpinnings for depressive symptoms. It is also reported that the social media posts from many patients in the clinical setting do not contain these signals \citep{ernala2019methodological}. Based on this research gap, we propose to monitor a type of signal that is well-established as a class of symptom in affective disorders --- mood. Mood is an experience of feeling that can last for hours, days or even weeks \citep{american2013diagnostic}.  In this work, we attempt to enrich current  technology for detecting symptoms of potential depression by constructing a 'mood profile' for social media users. 
%ADD DEFINITION TO WHAT YOU MEAN BY MOOD HERE

The variance of quality and intensity of mood and emotional reactions are referred to as "affective style" \citep{davidson1998affective}, which underlies one's risks of developing psychological disorders \citep{rottenberg2003emotion, akiskal1996temperamental}. Assessing affective style in everyday life is difficult in an experimental context because it requires a costly extended period of data collection. In contrast, social media data contains longitudinal information that reflect one's emotional reactions to stimuli. Therefore, it can provide researchers with an alternative lens to examine the affective style of an individual, based on the premise that approval is obtained from social media users, and data privacy is well-protected. 

Existing models for detecting symptoms of potential depression often include mood as a feature variable in the modeling process. However, there are a few methodological gaps in these models. First, most of them do not distinguish between mood and emotions. Emotion is a brief reaction to a specific stimulus, whereas mood has longer temporal duration \citep{morris2012mood}. Researchers using social media data to study mood or emotions often see a single post as reflecting mood \citep{bollen2011modeling, thelwall2011sentiment, celli2016mood}. However, a single social media post is likely to reflect a participant's emotions at the time rather than ongoing mood \cite{batson1992differentiating, rottenberg2005mood}. In this current work, we adopted the definition of mood from  The Diagnostic and Statistical Manual of Mental Disorders \cite{american2000diagnostic}: ``mood is the pervasive and sustained `emotional climate', and emotions are `fluctuating changes in emotional `weather' ''. We sought to determine whether temporal mood representation derived from social media text is associated with subsequent self-reported depressive symptoms, and if so, what are the best approaches to represent mood as a time dependent variable for future work?

Furthermore, a majority of models in this line of research often ignore the fact that affect is inherently time dependent. Only a few models have adopted temporal affective patterns \citep{reece2017forecasting, de2013predicting}. Most of these models also formulate the associations between affect and depressive symptoms based on the averaged affect~\citep{schwartz2013personality, chen2020inspecting}, but the transitioning from one affective state to another was largely ignored~\citep{rottenberg2005mood, frijda1993moods,bylsma2011emotional, sheppes2015emotion}. In this work, we explored and tested multiple approaches to represent the temporal affective patterns and the transitions of affective states. 

Nevertheless, social media users often post sporadically. The sparsity of social media data posits a big challenge in the modeling process. Most of the existing studies imputed  missing values with the mean or simply removed users with a lower word count \citep{de2013predicting, wang2013depression}. Removing outliers is beneficial to the modeling process. However, it may result in removing those with severe symptoms from the sample, because disinterest in social contact and social withdrawal (e.g. posting sparsely) is the core symptom of major depressive disorder (MDD)  \citep{american2000diagnostic}. Therefore, it is necessary to use some modeling techniques to include the outliers.

Towards addressing the methodological gaps described above, we designed multiple mood representations with the following characteristics: (i) Temporal features (ii) Transitions from one mood state to another (iii) Posting behavior. Here we see all the mood representations as a \textit{Mood Profile} for social media users. We formulate the following questions to explore the roles of mood in predicting depressive symptoms:

 \begin{enumerate}
 \item Are mood representations derived from social media text associated with the severity of self-reported depressive symptoms?
 \item Which representation in the mood profile is most predictive of the severity of self-reported depressive symptoms?
 %\item How precisely can we infer the degree of depressive symptoms using mood pattern derived from social media data?
 %In what way does mood indicated by social media text reflect depressive symptoms? (answer: fluctuation, transition)
 %\item Does silence behaviour have implications for predicting depressive symptoms?
% \item Does including dynamic social media mood as feature variables further improve the existing automatic symptom screening technology?
 \end{enumerate}

Our main contributions in this study are:

\begin{enumerate}
\item Constructing a mood profile for social media users based on their status updates. The mood profile encompasses representations that encoded the variance of mood intensity, alternations of mood states and the behavior of not posting.
\item Examining the associations between the social media mood profile and users' depressive symptoms level.
\item Examining which representation in the mood profile is more predictive to depressive symptoms level.
\end{enumerate}

In our work, we analysed a set of 93,378 posts from 781 Facebook users who had consented to the use of their posts and answers to related questionnaires for research reasons. For each user, a mood profile is constructed based on their social media text. We found that people with low symptom level tend to have less fluctuations in their mood pattern. We also modelled the mood representation with a Hidden Markov Model and we found the hidden states estimated based on the mood representation is highly related to depressive symptoms. Nevertheless, combining several representations in the mood profile is more predictive to depressive symptom levels (f-score: 0.62) than using one representation only. Our results suggest the mood profile derived from social media text can potentially serve as a reference for an individual's depressive symptom level. The data-driven, evidential nature of our approach provides us with better insight into the relationship between mood derived from social media data and depression.

\section{Background}
\subsection{Depression and Mood}
Moods are slow-moving states of feeling, influenced by others, objects or situations \citep{rottenberg2003emotion,watson2000mood}. The pattern of mood reflects one's vulnerability to developing affective disorders \citep{rottenberg2005mood, rottenberg2003emotion}. Depressed mood is a symptom of mood disorders, such as major depressive disorder (characterized by a persistent feeling of sadness) and dysthymia (persistent mild depression) \cite{american2013diagnostic}.

%For example, depressed mood often co-occur with a narrow-focused, preoccupying thinking patterns called rumination \cite{park2004effects, pronin2008thought}. 
%Clinical literature classify mood states into three types: depressive, manic and bipolar.

It is also well established that mood fluctuation and irritability are associated with many somatic and sensory dysfunctions in the psychology literature. Frequent alternating between moods (typically a few days) and irregular cycles of mood underlie the behavioural features of a wide variety of conditions \cite{akiskal1996temperamental}. In this study, we expect to find associations between mood derived from social media text and depressive symptoms similar to the psychology literature. Some level of associations has been found in the existing studies. For example, participants with depressive symptoms use more negative affective words (e.g. sad, cry, hate) in their social media text than those who do not \cite{de2013predicting,park2012depressive}. 
%It has been found that social media users that report high depressive symptoms tend to use more negative words in terms of ratio (with?) \citep{tsugawa2015recognizing, de2013predicting, park2012depressive, nguyen2014affective}.

\subsection{Detecting Depressive Symptoms with Sentiment}
Studies which examine emotions derived from social media data often adopt sentiment analysis. This is a computational process that categorizes affect or opinions expressed in a piece of text. The extracted affect is called sentiment \citep{pang2008opinion}. Most of the existing works use averaged sentiment over a long period of time (e.g. one year) as a feature to predict depressive symptoms \citep{coppersmith2014quantifying, tsugawa2015recognizing, benton2017multi, park2012depressive,tsugawa2015recognizing, wang2013depression}.

%Existing studies often use combations the ratio of positive/negative sentiment, average sentiment over a long period of time, together with analysis of topics in the text, n-gram and social network information to predict self-reported symptoms \cite{coppersmith2014quantifying, tsugawa2015recognizing, benton2017multi, park2012depressive,tsugawa2015recognizing, wang2013depression}. 

In addition to that, the change of sentiment over time is also an important aspect to infer affective disorders. However, only a few studies have included sentiment as a time dependent feature in the model \citep{de2013predicting}. For example, \citet{de2013predicting} used the momentum of the feature vector in the screening detection. \citet{eichstaedt2018facebook} include temporal posting patterns, but not the temporal affect pattern. \citet{chen2018mood} used temporal measures of fine grained emotions to predict users' depressive states. Recently, \citet{reece2017forecasting} adopted a Hidden Markov Model (HMM)  to analyse the change of language in social media posts and users' depressive symptom. They found that the shift of words in status updates indicate depression and (expand) PTSD symptoms. The above mentioned studies adopted a sliding window technique to define dynamic sentiment \citep{de2013predicting, chen2018mood, reece2017forecasting}. However, none of them systematically explored the size of time window and the slide increment, and most studies  only use a continuous sentiment value.  In this work, we  aggregated the sentiment in a sliding window based on its dominant valence (e.g. positive, negative) or average value. We also included the changes of affective states as a feature variable. 
%there is still a lack of studies that explore different approaches to include mood as a dynamic variable in computational models. In this study, we attempt to reflect mood as a dynamic feature, we defined the derived social media mood as the dominant emotion (sentiment) within a sliding time window to retrieve dynamic mood data. 

\subsection{Posting Behavior and depressive symptoms} 
Social media users are known to communicate selectively due to self-presentation biases \cite{kim2011facebook, vogel2014social}. They are less likely to reveal events that project negatively on themselves \citep{mehdizadeh2010self} due to stigma and fear of potential repercussions. Therefore, self-presentation biases leads to fundamental differences between real-life mood and social media mood.  

In addition to that, social media behavior can be counter intuitive. For example, people with who are more depresses would be expected to post less than people with fewer symptoms, however, several studies found that individuals with a history of depression (determined from past medical history) tended to post more often compared with people without depression \citep{smith2017variations}. There are several potential reasons for this. A person might not be severely depressed, they might be more comfortable with talking about their feelings, they might see their social media as a place where they can escape stigma, or they might have a social media support network for their mental health.  In this study, we see the behaviour of not posting as a variable in itself and observe if  posting frequency has any predictive capacity with regards depressive symptoms.

\section{Data}
For this study, the myPersonality data set~\citep{bachrach2012personality,youyou2015computer} was used. It contains Facebook posts of 180,000 participants collected from 2010 to 2012, enriched with a variety of additional validated scales \citep{bachrach2012personality}. The collection of myPersonality data complied with the terms of Facebook service, and informed consent for research use was obtained from all participants. Permission for the use of this database was obtained in 2018, and Ethical Approval for this piece of secondary data analysis was obtained from the Ethics Committee of the School of Informatics, University of Edinburgh. Other publications using this dataset include \citep{freudenstein2019four, sun2019group}.

\subsection{Screening for Depressive Symptoms}
From the participants in the myPersonality data, we focused on 1047 participants who completed the Center for Epidemiologic Studies Depression Scale (CES-D). The CES-D is a 20 item scale that measures the presence of depressive symptoms in the general population~\citep{Radloff77}. It is one of the screening tests most widely used by health service provider. The symptoms measured in CES-D include mood, anhedonia, the feeling of being worried, restless, changes in sleeping pattern and physical symptoms (such as lost of appetite) and irrational thoughts. The scale has been found to have high internal consistency, test-retest reliability \citep{Radloff77,Herz86,Roberts80}, and validity \citep{Herz86}. 

\citet{Radloff77} proposed three groups of depression severity: low (0-15), mild to moderate (16-22), and high  (23-60). For using mood profile to predict self-reported depressive symptoms, we followed the practice from previous social media studies \citep{de2013predicting,park2012depressive, reece2017forecasting, tsugawa2015recognizing} and adopted 22 as a cutoff point to divide participants into high symptoms and low symptom groups. This allows us to compare our model's performances with previous studies. For examining the mood fluctuation, we were additionally interested in a more nuanced picture in different symptom levels. Therefore, we further distinguish moderate and high symptom by
following the original study from \citet{Radloff77}. Participants were divided into three groups using two cutoff points: 16 and 22.

\subsection{Summary Statistics}
%Among 180,000 participants who allowed myPersonality to collect their Facebook posts, only 1047 (0.5\%) of them completed CES-D and allowed myPersonality to collect their demographic information. 
Among the 1047 participants who completed the CES-D scale, we removed 110 participants who were less than 18 years old. The CES-D survey was open from 2010 to December 2012, but MyPersonality only collected participants' status updates from January 2009 to December 2011. Since 2012 status updates were not available, we further removed participants who completed the scale in 2012 and who posted at least one post in the past year. Eventually we yielded a final set of 781 participants who had posted 93,378 posts over the past year before they took the test. 

\begin{figure}[hbtp]
\caption{Distribution of post count from participants}
\includegraphics[width=\columnwidth]{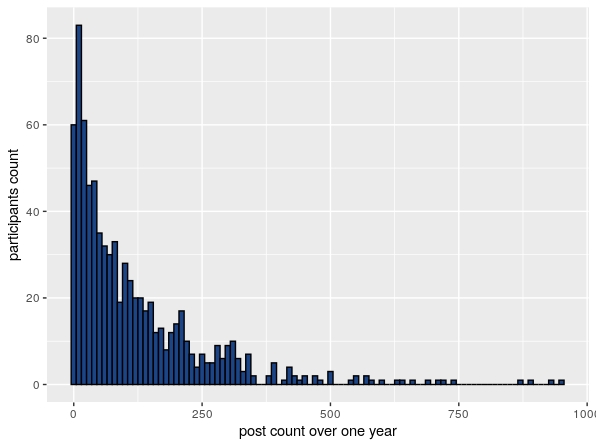}
%\fnote{Note: one two}
\footnotesize
\emph{Note: Figure demonstrates the distribution of post count over one year before participants completed the CES-D survey scale. Size of the bin is 10.}
\label{fig:freq}
\end{figure}

The average number of posts per user over one year was 120, this distribution was skewed by a small number of frequent posters, as evidenced by a median value of 73 posts per user. Figure~\ref{fig:freq} shows participants' count of posts up to one year before they completed the CES-D scale. The mean age of the participant is 26 ($sd$ = 11.7), 333 (43\%) participants are male and 448 (57\%) are female. Table \ref{tab:ethnic} shows further details of the participants, including the ethnicity, gender and marital status.

\begin{table}[hbtp]
\caption{Demographic Information of the 781 Participants}
\small
\begin{tabular}{|l|r|r||l|r|r|}

\hline
Ethnicity                            & No.   & \%   & Marital Status               & No.   & \%   \\
\hline
Black     & 38  & 4.3    & Single                         &574  & 73.8 \\
Asian Chinese                         &  26 & 3.3    & Divorced                       & 28  & 3.5  \\
Middle Eastern                      &  13    & 1.7         & Married                    &  27    & 3.4  \\
Native American                       &  13  & 1.6         & Married with Children                &  38  & 4    \\
Other Asians                        &   84     & 10.8      & Partner & 78 & 10   \\
Not Specified                          &   96  & 12.2       & Not specified                 & 36  & 4.5  \\
White-American                       &    309   & 39.2      &                               &   &      \\
White-British                          &    71  & 8.9      &                              &    &      \\
White-Other            &     131  & 17.1     &                              &   &  \\
\hline
\end{tabular}
\label{tab:ethnic}
\end{table}

\begin{figure}[hbtp]
\caption{Distribution of CES-D score}
\includegraphics[width=\columnwidth]{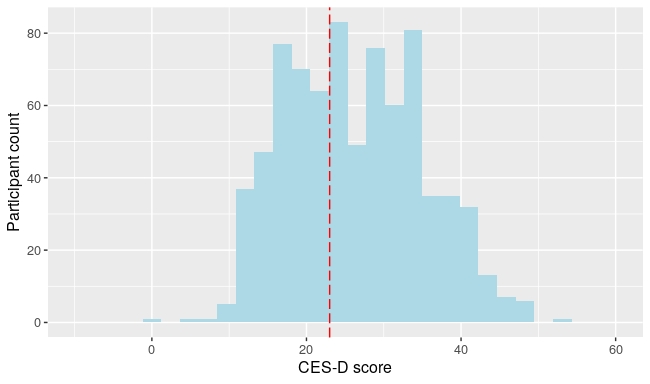}
%\fnote{Note: one two}
\footnotesize
\emph{Note: Figure demonstrates the density distribution of the CES-D score, red line indicate the cutoff point 22}
\label{fig:cesd}
\end{figure}

Overall, our sample has a relatively high mean CES-D score ($m$ = 26.3, $sd$ = 8.9), and the proportion of high symptom class to low symptom class is 1.6:1 (cutoff 22), see Figure \ref{fig:cesd}. \citet{radloff1977ces} found only 21\% of the general population scored at and above an arbitrary cutoff score of 16. However, we note the current dataset is not an exceptional case.  For example, \citet{reece2017forecasting} used a dataset that contained 105 depressed participants and 99 non-depressed participants, other studies have a proportion of high symptom to low symptom class as 2:3 \citep{de2013predicting, tsugawa2015recognizing, nadeem2016identifying}, 3:5 \citep{orabi2018deep}. All of these studies recruited a sample biased towards potentially high symptom individuals compared with empirical studies which selected participants in a random trail. We speculate that there is a bias in those individuals self-selecting for this type of research.

\section{Constructing Mood Profile}

A mood profile is constructed for each participant. Each mood profile encompasses sets of features which represent mood, the change of mood and the transition of mood states. Since mood is time dependent, we use a sliding window technique to construct the temporal features.  A window starts from day 0 (the day when users completed the CES-D scale) and moves backwards for up to one year. Choosing the size of a time window presents  a challenging question, how granular should a time window be?  \citet{de2013predicting} look at a user's tweets in a single day. \citet{reece2017forecasting} use both day and week as the time window because most of the participants did not generate enough daily content. In this paper, we define the size of the time window as measured by day $d\in D := \{3, 7, 14, 30\}$, see Table \ref{tab:notation} for the notations. The size of the slide increment determines how much information the two adjacent windows share. The slide increment is also measured by day $s\in S := \{3, 7, 14\}$. 
%\textcolor{red}{To be specific, the mood profile, the time window and the slide increment can be represented as follow: [($start_1$, $end_1$), ($start_2$, $end_2$), ...], $d = end_1 - start_1$, $s = start_2 - end_1$.}

%Since social media users do not post every day, we encoded the behavior of not posting as "Silence" and we defined four mood states: positive, negative, neutral and silence. Each representation in the mood profile is comprised of many time windows. We assigned the most frequent mood state or the average sentiment in each time window as mood. 

%Considering a lot of the machine learning models, except for deep learning approach, do not take the order of the features into account, granular feature such as mood in 1 day might not be beneficial to the models.

%We adopt a manual feature engineer approach because such an approach benefit from higher interpretability and the engineered features could potentially provide more information to health care providers. Hence, many recent works in this field adopted models with statistics framework \citep{reece2017forecasting, reece2017instagram, eichstaedt2018facebook}. 
%

Another challenge is to decide how far back do social media posts indicate symptom level. Earlier studies use data up to one year before participants completed the self-reported symptom measurement \citep{de2013predicting}, \citet{reece2017forecasting} found that symptoms can be predicted up to nine months before the official disclosure of the illness. In the current work, each representation in the mood profile was constructed with posts written up to one year before the participant completed the CES-D survey.

\paragraph{Sentiment Scores} 
We used the sentiment scores retrieved from SentiStrength \citep{thelwall2010sentiment}. SentiStrength extracts sentiment from the text based on a function that describes how well the words and phrases of the text match a predefined set of sentiment-related words. 
%SentiStrength has been validated by numerous studies \citep{nielsen2011new, vilares2015megaphone, guzman2014sentiment}.

\paragraph{Temporal Mood Representations} 
Since many social media users do not post every day, we encoded the behavior of not posting as "Silence" and we defined four mood states: positive, negative, neutral and silence.  We adopted two approaches to define mood within a time window: most frequent mood state over a time window and average sentiment over a time window, see Table \ref{tab:notation}. If two mood states had the same high frequency in the same time window, we defined the mood as mixed. Since neutral mood state is relatively less frequent in compare with the rest of the mood states, we tend to give neutral more weights. If other mood states have the same frequencies as neutral, we defined the mood as neutral. For the average sentiment, silence days as missing values are imputed by the mean. We also constructed features that represent the change of mood  \citep{de2013predicting}, see mood momentum in Table \ref{tab:notation}.

\begin{table}[hbtp]
\caption{Notations for Mood Profile}
\label{tab:notation}
\small
\begin{tabular}{p{2.2cm}p{1cm}p{4.5cm}}
\toprule
Variable        & Notation    & Description                                                                     \\
\midrule
Window Size     & $d$         & A period of time within $d\in D$ days, $d\in D := \{3, 7, 14, 30\}$ \\
Slide Increment               &   $s$          &   A sliding window move forward by every $s\in S$ days, $s\in S := \{3, 7, 14\}$     \\
Sentiment & $v$ & Sentiment score of a single post \\
Day Sentiment & $V$ & Arithmetic mean of sentiment in one day $ V  = \dfrac{v + ... v_i}{i}$\\
$Mood_\mu$      & $M_\mu$     & Arithmetic mean of day sentiment over a time window, $ M_\mu =\dfrac{V + ... V_d}{d}$                                           \\
$Mood_\omega$   & $M_\omega$  & Most frequent sentiment over a time window, categorical                         \\
Mood Momentum   & $\Delta M $ & Difference between $M_\mu$ in two time windows                                  \\
Mood States Transition & $Tr $       &  The probability of a user transfer from one mood-state to another, a mood state is defined by  $M_\omega$  
\\
Mood States Transition & $\Delta Tr $       &  Difference between $Tr $ in two time windows                          \\
% Silence     & $\emptyset$        & User did not post any content in a time window                                  \\

\midrule
\end{tabular}
\end{table}

\paragraph{Temporal Mood Transition Representations} 
We also encoded the probability of a user transferring from one mood state to another as a representation in the mood profile. We have in total 16 transition states (e.g. positive to negative, negative to silence) from the fours classes (positive, negative, neutral and silence). Note that if we set the slide increment as one day, we would have 365 $\times$ 16 mood transitions features. To prevent the large dimensionality, which might led to sparse representation, we defined $d$ as 30 and $s$ as 30, so that we have  12 $\times$ 16 feature columns for Mood Transition Representations.

\section{Association Between Mood Profile and Depressive Symptoms}
We first observed whether the pattern of the mood profile is related to symptom level. Then we tested the mood profile's predictive power on symptom level.

\subsection{Mood Fluctuations}
\begin{figure}[hbtp]
\caption{Example of GP Regression}
\includegraphics[width=85mm, scale = 1]{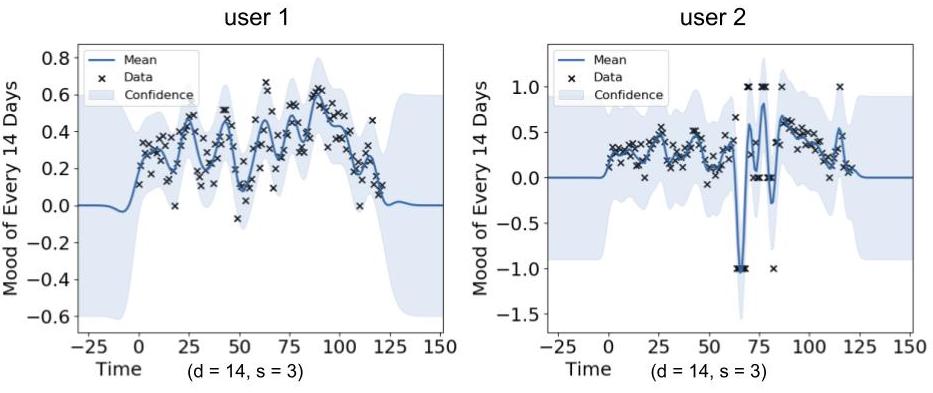}
%\fnote{Note: one two}
\footnotesize
\emph{Note: here shows examples from two participants, each data point represents mood of every 14 days estimated by the GP regression model. $N$ = 690.}
\label{fig:gp}
\end{figure}

We modelled mood fluctuations using Gaussian Process (GP) regression. GP regression is a Bayesian approach that assumes a Gaussian process prior over functions \citep{quinonero2005unifying}.  In this analysis, we see the temporal mood representations as noisy representations of participants' mood. We use GP regression to estimate participants' latent mood based on their mood representation. For participants with few data points, the GP regression is modeling the mean of the sample due to the imputation approach we adopted. Thus, for this experiment, we excluded participants posted less than 10 posts over year before they completed the depressive symptom scale. Eventually, this yielded 690 participants for the current analysis. We used mood representations with $d\in D := \{1, 3, 7, 14\}$ and  $s\in S := \{1,3,7\}$ as input of the GP regression model. The GP regression is best fitted on mood vector with $d$ = 14 and $s$ = 3, see Figure~\ref{fig:gp}. Each dot on the graph represents mood (averaged sentiment) in a time window $d$ = 14, x axis shows the count of time windows. Since the entirety of the dataset includes posts of one year (365 days), there are 122 time windows for each participant.

We constructed one model for each participant. Here we are not interested in making prediction with the GP regression model, instead, our focus is on the function parameter, lengthscale. The lengthscale describes how smooth a function it is. A small lengthscale means the function value changes quickly, while a large lengthscale means that its value changes slowly  \citep{chalupka2013framework}. By fitting a GP regression model on each user, we obtain a lengthscale of each user's latent mood, and we compare the lengthscale among participants with different symptom levels (low, moderate, high).

% \rule{1cm}{1cm}% placeholder for `\includegraphics`
% \caption{A figure}
% \floatfoot{A note}
% \end{figure}

We used a nonparametric test (Mann-Whitney U test) to compare the lengthscale differences between groups. The lengthscale of the high symptom group ($Median$ = 2.77) is identical to the moderate symptom ($Median$ = 2.77) group ($U$ = 35424, $p$ = 0.01).  However, the low symptom group  ($Median$ = 2.98)  has a significantly larger lengthscale than the high symptom group ($U$ = 17231, $p$ = 0.01). The moderate symptom group was also significantly different from the low symptom group ($U$ = 7244, $p$ = 0.02). Our result suggest that people with high or moderate depressive symptom level have more mood fluctuations than people with low symptom level.

\subsection{Classifying Symptom Levels using Daily Mood Representation}
Another approach to examine whether the mood profile is associated with depressive symptom is to see if a particular mood state is influenced by depressive symptoms level. We assume the mood states are serially dependent and we used Hidden Markov Model (HMM) \citep{beal2002infinite} to model two unobservable states based on a daily mood state representation. This representation comprises four mood states (positive, negative, neutral and silence). Since the behavior of not posting (silence) is included in the modeling process, we did not remove any less active users in this analysis (N = 781).

\paragraph{Hidden Markov Model}
We used a multinomial (discrete) emission Hidden Markov Model (HMM) to model users' observed mood for one year \citep{johansson2007bayesian}. The major parameters used for the model are:

\begin{enumerate}
    \item Observed mood $O_t$ (time series), daily mood transition representation ($d$ = 1, $s$ = 1).
    \item Transition matrix ($A$), gives the probability of a transition from one state to another.
    \item Transition state $j$.
    \item Observation emission matrix ($B$), which gives the probability of observing $O_t$ when in state $j$.
\end{enumerate}

An HMM model (denoted by $\lambda$) can be written as:

\begin{equation} \label{eq:8}
\lambda = (\pi, A, B)
\end{equation}

The idea behind this approach is to use the observed mood to estimate the parameter set $(\pi, A, B)$, $A$ shows us the probability of transferring from one hidden state to another, and $B$ tells us the probability of emitting a certain mood when a user is in a specific symptom state.

We used hmmlearn python library \citep{gao2017complex} to fit emission, transition matrices (using expectation-maximization) and hidden state sequence (using the Viterbi path algorithm), see Section \ref{appendix} for the initialized probabilities. We trained the model on the entire set of data and observed if the emission probabilities align with our existing knowledge of affect and depressive symptoms. Here we were not to find the optimal model to forecast a new observation sequence, hence we did not test the training model on a test set. Instead, we were interested to know whether the hidden states decoded from the HMM model were associated with depressive symptom levels. 

The HMM model decodes a binary hidden state for each day. We speculate that one of the hidden states represents the user experiencing more depressive symptoms (high symptom state), and another represents fewer symptoms (low symptom state). Although the CES-D scale measures an overall symptom level in one week, it is entirely possible for an individual to have more symptom on some days (e.g.sleep disturbance, loss of appetite) but less on others.  To test our speculation on the hidden states, we classify participants' self-reported symptom level according to the count of high symptom state. Here we use cutoff score 22 to divide participants into two groups for comparing the results with the existing models. However, there is a challenging questions, up till when shall we count the high symptom states?  Since the CES-D scale measures an overall symptom level in the past one week  (e.g less than 1, 1-2, 3-4, 5 or more), and the Diagnostic and Statistical Manual of Mental Disorders, 5th Edition (DSM-5) defines depressive symptom as ‘‘The individual must be experiencing five or more symptoms during the same 2-week period''.  Therefore, we defined our classification criteria as whether participants have at least $x$ days experiencing high symptom in the last $y$ days before they completed the CES-D scale, $x\in X := \{1,2,3,4,5,6,7\}$, $y\in Y := \{7,14\}$.

\subsubsection{Evaluation of Hidden States}
\paragraph{Emission Probabilities}
We observed whether the hidden states' emission probabilities align with our existing knowledge in depressive symptom and affect. Table \ref{tab:emission} shows two hidden states and their emission probabilities to each observation. Given an observed day, we can see both hidden states were most likely to emit silence day because social media users posted sparsely. However, the high symptom hidden state has lower probability to emit silence days compared with low symptom hidden state. The high symptoms state also has a higher probability to emit negative mood or neutral mood, but the low symptoms state has a higher probability to emit positive mood. Therefore, results from the HMM model aligns with our existing knowledge in depressive symptom and affect.

\begin{table}[hbtp]
\centering
\caption{Emission Probabilities}
\label{tab:emission}
\begin{tabular}{lllll}

                  \hline
   $N = 781$               & Positive     & Negative   & Neutral   & Silence  \\
                  \hline
Low Symptom  & 8.51         & 5.20       & 4.65      & 81.6     \\
\hline
High Symptom   & 3.15         & 12.8       & 7.00      & 76.9     \\
\hline

\end{tabular}
%  \begin{tablenotes}
%       \small
%       \item Note: less symptoms: hidden state that represents less symptoms on a particular day, more symptoms: hidden state that represents more symptoms on a particular day, $N$: training sample size
%     \end{tablenotes}
    
 {\raggedright Note: less symptoms: hidden state that represents less symptoms on a particular day, more symptoms: hidden state that represents more symptoms on a particular day, $N$: training sample size \par}
\end{table}

\paragraph{Transition Probabilities of Observations}
We are also interested to know whether people are more likely to transfer from certain mood states to another. We constructed a transition probability matrix for the observations (daily mood representation). Table \ref{tab:tranPro} again shows us that social media users in general are more likely to become silent after they posted any social media content, although high symptom group is less so. High symptom individuals have higher probabilities of changing in between any mood states other than silence. This result aligns with the findings from the GP regression that low symptom individuals shows less fluctuations in their mood representation.

In general, people were more likely to have a positive mood if they had a positive mood in the previous time window. The probabilities of $+ \rightarrow +$, $- \rightarrow -$ were similar among the two groups, but high symptom participants are slightly more likely to transfer from negative to negative. When low symptom participants have a neutral mood, they have similar chances of having a neutral or negative mood in the next time window, whereas, high symptom participants are also more likely to have a negative mood in the next time window. Our result shows that while people, in general, are more vocal when they have a negative mood, but high symptom participants are more likely to vocal about the negative content for a more extended period. 
%With these results, we encourage further analysis of the content nature and its relationship with posting behaviour.

\begin{table}[hbtp]
\centering
\caption{Transition Probabilities of Observations}
\label{tab:tranPro}
\begin{tabular}{llllllllll}
\hline
  & \multicolumn{4}{l}{High symptom} &  & \multicolumn{4}{l}{Low symptom} \\
  \hline
  & +       & -      & 0     & S     &  & +      & -      & 0     & S     \\\hline
+ & 21.1    & 15.7   & 13.4  & 49.6  &  & 19.5   & 13.3   & 12.3  & 54.8  \\
- & 22.3    & 16.2   & 14.1  & 47.3  &  & 20.5   & 13.3   & 12.9  & 53.3  \\
0 & 19.3   & 14.5   & 12.8  & 53.3  &  & 17.6   & 11.6   & 11.8  & 58.9  \\
S & 5.82    & 37.5   & 4.21  & 86.2  &  & 5.92   & 37.1   & 4.33  & 85.9 \\
\hline
\end{tabular}
%  \begin{tablenotes}
%       \small
%       \item Note: $+$: positive, $-$: negative, $0$ neutral, $S$: silent
%     \end{tablenotes}
{\raggedright Note: $+$: positive, $-$: negative, $0$ neutral, $S$: silent \par}
\end{table}

\begin{figure}[hbtp]
\caption{Precision and Recall of High Symptom Class (HMM) with Various Assignment Criterion}
\includegraphics[width=90mm, scale = 1]{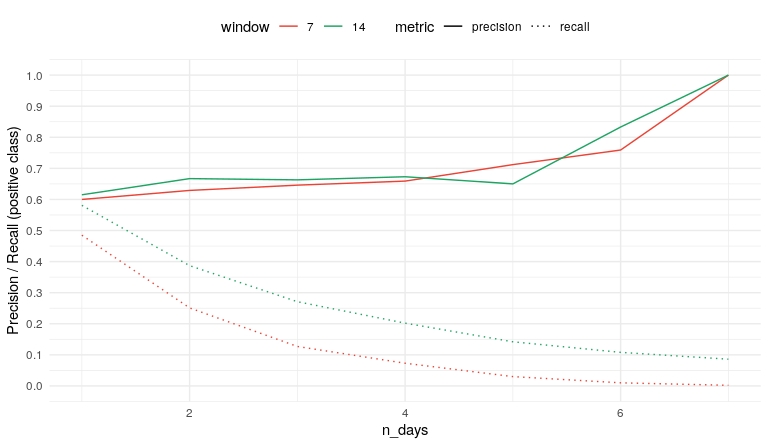}
%\fnote{Note: one two}
\footnotesize
\emph{Note: window: size of the time window, $x$ days before participants completed the CESD scale. ndays: count of high symptom state within the time window}
\label{fig:hmm}
\end{figure}

\paragraph{Using Hidden States to Classify Symptom Level}
Figure \ref{fig:hmm} shows the precision and recall of the high symptom class by counting the hidden states from the HMM model. The baseline model is formulated using a stratified dummy classifier that predicts based on the most frequent training labels. Precision increases as the criterion of $x$ increase. Table \ref{tab:hmmPre} shows some of the best classification results. Assigning participants with six high symptom states within 14 days to the high symptom class results in very low recall (10.8\%) but high precision (71.2\%). Assigning participants with one high symptom state within 14 days results in a more balanced recall (60.3\%) and precision (58.1\%)  to high symptom class. Result from this classifier does not surpass the baseline in f1 score but when using a higher $x$ as criteria, the precision rate is much higher than the baseline. Our result supports the claim that daily mood representations inferred from social media text is highly associated with depressive symptoms. When a social media user shows specific mood patterns, it is highly likely that the person developed high level of depressive symptoms. However, only using this approach to identify high symptom individuals would result in a lot of false negative cases.  

\begin{table}[hbtp]
\caption{Predicting depressive symptom with hidden states}
\label{tab:hmmPre}
\begin{tabular}{llll}
\hline
 Criteria &P    & R     & f1      \\ 
 \hline
baseline & 61.2 & 100.0 & 76.0\\
$x$ = 1, $y$ = 7 & 61.5      & 48.5          & 45.2 \\
$x$ = 1, $y$ = 14  & 60.3       & 58.1          &59.2 \\
$x$ = 6, $y$ = 14  & 71.2      & 10.8         &18.9 \\
 \hline

\end{tabular}
%  \begin{tablenotes}
%       \small
%       \item Note: high: high symptom class, low: low symptom class, R: recall of high symptom class, P: precision of high symptom class, f1: average macro-f1 score of both classes, criteria: criteria for classifying high symptom class
%     \end{tablenotes}
    
{\raggedright Note: high: high symptom class, low: low symptom class, R: recall of high symptom class, P: precision of high symptom class, f1: average macro-f1 score of both classes, criteria: criteria for classifying high symptom class \par}
\end{table}

%We experimented $x$ with 1, 2, 3 days. We ran the trained model on a separate test set (30\% of the data), and we tested our criterion on the hidden states vector from the test set.
\section{Representation Predictability of Depressive symptoms}
%Finally, we tested the temporal mood representation sets in a logistic regression model to see if engineering different types of mood patterns can potentially improve the screening technology for depressive symptoms. 

The previous analysis suggests that the mood profile is highly associated with depressive symptoms. Now we examine which representation in the mood profile is most predictive of depressive symptoms. We combine the representations with sets of proxy signals in a classification task. 

\subsection{Feature Extractionn}
We extracted multiple features for the posts of each user to train multiple models for high-symptoms prediction. Our extracted featured included: 1) n-gram word representation, where $n\in N := \{1,2,3\}$); 2) topic modelling from Latent Dirichlet allocation (LDA) and 3) all the entries from Linguistic Inquiry and Word Count (LIWC) \citep{pennebaker2001linguistic}.
N-gram were ordered by term frequency across the corpus, we grid searched the number of most frequent n-gram and number of topics for LDA (see Section \ref{appendix}). We found the most frequent 1500 n-grams and 30 LDA topics gave us the optimal results. These feature variables were commonly used in detecting signs of potential depression \citep{de2013predicting,park2012depressive,coppersmith2014quantifying,reece2017forecasting}. We compare the precision and recall between models with different representations from the mood profile. 

Our dataset has an exceptionally high proportion of high symptom individuals as discussed earlier. Given that we have only 303 low symptom participants among 781 participants, we randomly selected 303 participants in the high symptom sample to have a dataset with a balanced class proportion that is closer to the existing literature (1:1), $N$ = 606, So that we can have results that are more comparable with the existing literature. We split the data into train (80\%, N = 486), and test set (20\%, N = 120) in stratified fashion. Stratified five-fold cross-validation was used to optimize the parameters in the model training. A grid search of parameters was carried out for several candidate classification algorithms (e.g. decision trees, support vector machine, logistics regression) \cite{suykens1999least}, see Section \ref{appendix} for the grid search parameters. 

\subsection{Model Evaluation}
\begin{figure}[hbtp]
\caption{Precision and Recall of logistic regression}
\includegraphics[width=90mm, scale = 1]{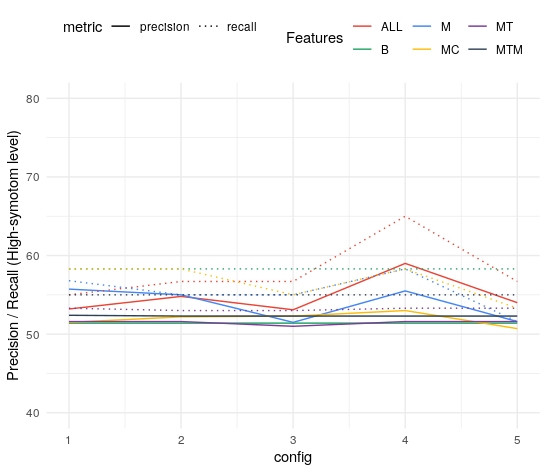}
%\fnote{Note: one two}
\footnotesize
\emph{Note: config 1: $d$ = 7, $s$ = 3, config 2: $d$ = 14, $s$ = 3, config 3: $d$ = 14, $s$ = 7, config 4: $d$ = 30, $s$ = 3, config 5: $d$ = 30, $s$ = 7, B: basic features (n-gram, topic modeling, LIWC), M: B + $Mood_\mu$, MC: B +  mood momentum, MT: B + mood transition, MTM: B + mood transition momentum, All: all features excluded MTM}
\label{fig:logistic regression}
\end{figure}

A baseline model is formulated using a stratified dummy classifier that generates predictions according to the training set's class distribution. Out of several candidate algorithms, logistic regression demonstrated best performance.  Hundreds of classification models were trained and evaluated for this task. The models with different representations from the mood profile can be evaluated by precision and recall. We grid searched $d$ and $s$ that maximises the metrics. Figure \ref{fig:logistic regression} shows the precision and recall of the high symptom class from models with various configurations and feature sets. Models with configuration 4 (time window 30 days and increment slide 3 days) yield the best scores. Table \ref{tab:pred} shows the precision, recall and f1 score of the high symptom class from configuration 4. The model with mood, mood momentum and mood transition representations yields the highest scores, and the model with averaged mood over a time window gives second highest scores, 0.59 precision, 0.65 recall, and an F-score of 0.62. 

%Models with mood transition representation yields a high precision at the cost of low recall. This behavior is consistent with using hidden states from the HMM to classify symptom level. The HMM is also modeling the mood transition states. Models using all the representations in the mood profile gives better recall at the cost of precision. Mood momentum feature gives a relatively balanced precision and recall in two configurations.

%Table \ref{tab:pred} shows the precision and recall of the high symptom class, and the averaged f1 score of the positive class when $d$ = 7, $s$ = 3. We observed a slight improvement of the result by adding representations from the mood profile to the proxy signals. Momentum features have a more significant improvement for the result probably because they capture the changes of the variable. However, using mood profile in a time-series approach to classify symptom level achieved similar precision rate as using combined sets of proxy signals in an logistic regression model. Our result suggests an analysis of more sophisticated time-series models would be an essential topic for future work. 

\begin{table}[hbtp]
\caption{Prediction result of depressive symptom ($d$=14, $s$=3}
\label{tab:pred}
\begin{tabular}{l|lll}
\toprule
Features    & P  & R   & F1  \\\hline

RB & 47.6 & 50.0 & 48.8 \\
B                            & 51.4                &58.3 & 54.7     \\
B + $M_\mu$                        & 55.5                & 58.3 & 56.9       \\

%Pr + $M_\omega$ & 61.8 &  55.8 &  58.4    \\
B + $\Delta m$  &53.0              & 58.3 & 55.6     \\
B + $Tr$                       & 51.6             & 53.3 & 52.4        \\
B + $\Delta Tr$                       & 52.3           & 55 & 53.6        \\
B +  $M_\mu$+  $\Delta m$ +   B + $Tr$       & \textbf{59.0}              & \textbf{65} & \textbf{61.9} \\\toprule

\end{tabular}
%  \begin{tablenotes}
%       \small
%       \item Note: R, P, F1 are recall, precision, and f1 score of high symptom classes respectively. B: basic features (tfidf bag-of-words, topic modeling, sentiment, LIWC). RB: random baseline, model parameters: penalty: l2, Inverse of regularization strength: 0.1)
%     \end{tablenotes}
    
{\raggedright Note: R, P, F1 are recall, precision, and f1 score of high symptom classes respectively. B: basic features (tfidf bag-of-words, topic modeling, sentiment, LIWC). RB: random baseline, model parameters: penalty: l2, Inverse of regularization strength: 0.1  \par}
\end{table}

\section{Discussion}
\subsection{The Role of Mood in Predicting Depressive symptoms}
%The present study aims to identify the mood pattern of social media users and see its implications for depressive symptoms. 
%Mood is an important predictive marker of affective disorders, monitoring the pattern of mood provides us insights on a variety of affective disorders other than just depression. Even though mood inferred from social media data is different from daily life mood, 

Mood is a time dependent variable, using time series approaches to model mood inferred from social media text provides us with better insight about mood and depressive symptoms. Participants in this study demonstrated significantly fewer mood fluctuations if they reported a low symptom score. This finding aligns with the well-established connection between emotionality and depression in the psychology literature. We also found the hidden states from the HMM model are highly relevant to self-reported depressive symptoms, see Table \ref{tab:hmmPre}. Our model suggests that an individual having  one high symptom state in 14 days is highly likely to have high symptom level. It is worth to note that the criteria we used in here is different from the criteria in the CES-D scale, where individuals need to have  experienced symptoms 1-2 days in the past 7 days to score on a criterion. However, we cannot assume that people will talk about their symptoms every time they experience them. This result suggests that individuals who show specific mood pattern in social media text are highly likely having high depressive symptoms, however, most of the individuals with high symptom do not display this mood pattern. 

Existing studies that use a sliding window technique to create dynamic sentiment features have not yet explored which representations and configurations tend to yield a better result in classifying symptoms. We explored various configurations of the sliding window and found that mood in a 30 days time window and move the time window every 3 days is most predictive to depressive symptom level. This result suggests that a less granular mood representation is more beneficial in identifying symptoms. Moreover, combining several representations in the mood profile together can dramatically enhance the model performance. Our best model (f-score: 0.62) encompasses the mood profile and a set of basic features commonly used in existing works. Other studies using multiple sets of proxy signals to predict depressive symptoms achieved a precision score ranging from 0.48 \citep{coppersmith2014quantifying} to 0.87 \citep{reece2017forecasting, guntuku2017detecting}. \citet{schwartz2014towards}, using the same data set, achieved correlation of 0.386 with  continuous scores. The mood profile can potentially enhance the current screening technology by combining it with more advanced engineered features.

%This result suggests that mood pattern derived from social media text is associated with one's emotionality. However, the magnitude of the association varies from person to person. Health care providers can leverage this technique to access an abstract representation of people's trait-like affective pattern. 
%This intriguing pattern suggests that some participants might show signs of rumination, as evident in a recent study \citep{chen2020inspecting}. 

The transition probabilities of mood showed that participants, in general, were more vocal on social media when they were in a negative mood. %This seemed to be contradicted with the fact that people with depression suffer from a loss of pleasure in usual activities. 
We speculate that some depressed individuals react to negative mood by posting, and some by silence. 
Those who are more vocal on social media when in a negative mood might be using social media to reach out to others or use posts as a way to reflect.  %This sometimes were interpreted as social media platform making people depressed \citep{pantic2014online}. 
The associations between negative mood and being vocal, and the association between high symptom scores and a specific mood pattern, suggest that posters could be stratified into several groups, those that withdraw, those that reach out, and those that do not disclose potential signs of depression on social media.

Our results show that a temporal mood profile derived from social media text is highly associated with users' subsequent self reported depressive symptom level. In order to examine the potential of mood momentum and mood transition further,  advanced time series analysis techniques need to be applied. Most importantly, mood profiles can potentially provide more information to clinicians than a classification system with binary output. 

\subsection{Technological and Ethical Implications}
%A recent study suggests that incorporating clinical judgment via an appraisal of social media self-reports of mental illnesses leads to the best performance \citep{ernala2019methodological}. However, this finding only relevant for those people who disclose or discuss their condition on Social Media. 

Similar to the existing studies, the present finding of the derived mood pattern has implications on symptom level but does not provide an accurate interpretation for participants' mental health condition. An accurate interpretation of one's mental health condition requires a holistic view, and any diagnosis requires a strong understanding of an individual's case history. The daily life information contained in social media data is just a tip of the iceberg of one's life experience.
%Therefore, the desired outcome of using this technique should be to identify individuals with mood abnormality that might expose one to a certain level of vulnerability to develop affective symptoms. Participants need to be informed of their scores and a diagnosis result should be a collaborative work from both the patient and providers. 
%Whether a person is evaluated for a particular disorder should be based on health care providers' judgement. 

Our approaches provide a useful source of information for assessing participants’ derived mood pattern over time. However, as with all social media related research, ethical and privacy issues need to be considered, given the potential for misusing social media data \cite{fiske2014protecting, lumb2016scientists,cadwalladr2018revealed}. Using social media analysis techniques in practice requires that the user whose data is being analysed is comfortable with their social media timeline being used in this way, and that they consent to it. The scope of their consent also needs to be clear, i.e., whether it is for research or whether it is also for potential clinical use.

\subsection{Limitations}
Our sample contains participants who allowed researchers access to their Facebook posts and to complete a symptom screening scale. Therefore, this sample may be strongly biased towards those who were comfortable to disclose and reach out on social media. It is still unclear about what the biases are in a sample with these tendencies compared with a random patient sample. Of particular interest is the relatively high depressive symptom score from most of the participants in this sample, and this bias is prevalent in studies in this line of research \citep{guntuku2017detecting}. We speculate that people who have depression are more curious about taking part in mental health related studies.

In this work, the symptom screening test was conducted once only. There were also no tests controlling for the presence of other disorders, such as bipolar, which greatly affect behaviour and mood variability. Those at the high end of the scale could have other types of affective disorders but showing depressive symptoms at the time when they carried out the self-reported measurement. Therefore, the measurement of self-reported symptoms is not an accurate reflection of whether the person has depression.

In addition, the sentiment scores employed in this study were retrieved with SentiStrength, which is a word counting approach to identify positive and negative affect. Although numerous studies have validated the word counting approach, the ideal method to retrieve less noisy sentiment is to construct the sentiment classification model with the examined dataset. Future studies can train their model for sentiment annotation to retrieve more accurate sentiment.

\section{Conclusion}
Mood is an important signal for the development of a depression episode. This report provides an outline of utilizing the sliding window technique to construct temporal representations of mood based on sentiment expressed in social media text. The behavior of not posting was also encoded in some of the representations. However, mood inferred from social media text is different from mood in real life. In order to examine whether the mood profile inferred from social media text is associated with depressive symptoms, we use the mood profile to classify depressive symptom level with time-series modelling and logistic regression algorithm. Our result suggests that the mood profile inferred from social media data is highly predictive of depressive symptoms, especially when the behavior of not posting is included. We also discover a pattern whereby people are more vocal in social media when they are unhappy. Despite many social media users being subject to positive self-presentation biases, social media provides a place for people to channel their emotions. Future studies can focus on studying this behavior on an actual patient group and a random control sample. The techniques proposed here offer a novel contribution to technology for detecting potential signs of depression as they are not focused on providing a binary classification result, but a longitudinal reference for the development of depressive symptoms.

%Our findings  support the claim that derived mood from social media text can be used as a proxy of real-life mood to infer depressive symptoms in the current sample.

%The resulting models can be used as an inexpensive means for affective symptom analysis. Our results support that mood derived from social media data is highly predictive of self-reported depressive symptoms, even when it is not combined with any other proxy signals. 

\begin{acks}
We thank Michael Kosinski and David Stilwell for permission to use myPersonality, the three anonymous reviewers for their useful comments, and Chris Lucas for invaluable guidance on Gaussian Processes. Magdy and Wolters acknowledge partial funding  by The Alan Turing Institute (EPSRC, EP/N510129/1).
\end{acks}

\appendix
\section{EXPERIMENT DETAILS}
\label{appendix}

The following supplementary material details what is required to reproduce our results as closely as possible. 
%The experiments used in running the pipeline to generate our results are defined in YAML files that are also provided on the \href{https://github.com/alan-turing-institute/DSSG19-HomelessLink-PUBLIC}{aforementioned public GitHub repository}. The repository also includes a more in-depth description of running the pipeline and the requirements of a system to do so.

\subsection{MODEL TRAINING} \label{appendix:model}

Grid searches of the following pairings of parameter spaces and Scikit-Learn implementations of algorithms were carried out:

\begin{itemize}
    \item Feature Extraction 
     \begin{itemize}
        \item  number of n-gram: 1000, 1500, 3000, 4000, 5000, 6000
        \item number of topics: 10, 20, 30
    \end{itemize}

    \item HMM:
     \begin{itemize}
        \item Initial transition probability:  [0.5, 0.5], [0.5, 0.5]
        \item Initial transition probability: [0.2, 0.3, 0,2, 0.3],
          [0.2, 0.2, 0.3, 0.3]    
         \item Number of iteration: 10
    \end{itemize}
    
    \item Support Vector Machine
    \begin{itemize}
        \item Inverse of regularization strength: 0.5, 0.7, 1.0, 1.5, 2.0, 2.5
        \item Kernel: linear, poly, rbf, sigmoid
        \item Kernel coefficient: 0.01, 0.001, 0.0005
    \end{itemize}
     \item Extra Trees 
    \begin{itemize}
        \item Number of Estimators: 100, 300, 500, 1000
        \item Maximum Tree Depth: 20, 50, 100, 200
        \item Maximum number of features: sqrt, log2
    \end{itemize}
    \item Logistic Regression:
    \begin{itemize}
        \item Penalty: l1, l2
        \item Inverse of regularization strength: 0.1, 0.3, 0.5, 0.7, 0.9, 1.0, 1.5, 2.0
    \end{itemize}

\end{itemize}

\bibliographystyle{ACM-Reference-Format}
\bibliography{citation}

\end{document}